\documentclass{PoS}

\usepackage{graphicx, epsf}
\usepackage{epsfig}
\usepackage{amssymb,amsmath, amsfonts}
\usepackage{bbold}

\def\be{\begin{equation}}
\def \ee {\end{equation}}
\newcommand{\SU}{{\rm SU}}

\newcommand{ \Real } {\mathrm{Re}}
\newcommand{ \Tr} {\mathrm{Tr}}

\newcommand{\ba}{\[\begin{aligned}}
\newcommand{\ea}{\end{aligned}\]}

\title{Five-dimensional Gauge Theories in a warped background}

\ShortTitle{Five-dimensional Gauge Theories in a warped background}

\author{Richard D. Kenway, \speaker{Eliana Lambrou} \\
	The Higgs Centre for Theoretical Physics, School of Physics and Astronomy,\\
        University of Edinburgh\\
        Edinburgh, EH9 3FD, UK\\
        E-mail: \email{r.d.kenway@ed.ac.uk, e.lambrou@ed.ac.uk}}

\abstract{The phase diagram of five-dimensional anisotropic gauge theories in a flat background has been extensively explored during the last decade. Here, we present novel results for the phase structure of the five-dimensional anisotropic \SU(2) model embedded in a warped background. The static potential in the deconfining region of the phase diagram, close to the transition to the layered phase, provides evidence of a Yukawa mass, suggesting that the system is in a 4D Higgs-like phase. As no symmetry has been broken by the boundary conditions, this phase appears to be due to the warp factor. Whether the system is dimensionally reduced from a 5D phase to this 4D Higgs-like phase, which would provide a mechanism for dimensional reduction via localization, remains open. 
\begin{flushright}
Edinburgh 2015/23\\
\end{flushright}
}

\FullConference{The 33rd International Symposium on Lattice Field Theory\\
		14 -18 July 2015\\
		Kobe International Conference Center, Kobe, Japan}

\begin{document}
\section{Introduction}
Extra-dimensional theories offer a solution to the hierarchy problem. Even though collider experiments have not provided evidence of the existence of extra dimensions, they have not excluded them either. All higher-dimensional theories must undergo dimensional reduction to be compatible with the observed four-dimensional world. This can be achieved by compactification, or localization and our work focuses on a possible way of achieving the latter. 
\par
Higher-dimensional theories are perturbatively non-renormalizable and therefore techniques of lattice gauge theories provide a tool for their investigation. Usually, phase diagrams are obtained and one seeks regions where the system is dimensionally reduced and a continuum theory can be defined by means of a second order phase transition. The phase diagram of the pure five-dimensional SU(2) lattice gauge theory has two phases (the 5D deconfining and confining phases), which are separated by a first-order phase transition and thus it is physically uninteresting.
\par   
In 1984, Fu and Nielsen showed that if an abelian anisotropic higher-dimensional lattice gauge theory is considered, there is an additional phase, called the layered phase~\cite{Fu}. In this new phase, the 4D hyperplanes transverse to the extra dimensions can be seen as layers where the gauge fields are localized. The existence of a critical point in the non-abelian case, where a 4D continuum theory could be defined is still in doubt~\cite{DelDebbio:2013rka}.  However, results from~\cite{MeanField1} suggest that interesting physics happens close to the transition line from the Coulomb to the layered phase.
\par
A well-known class of five-dimensional models are the so-called Randall-Sundrum models~\cite{Randall:1999ee, Randall:1999vf}, which are embedded in a warped background, given by 
\begin{equation}\label{eq:warped_metric}
ds^2 = \mbox{e}^{-2k|y|} \eta_{\mu\nu}dx^\mu dx^\nu + dy^2
\end{equation}
where $k$ is the curvature. We also define $f(y) = \mbox{e}^{-2k|y|}$, that is called the warp factor, for later use. In these models, the extra dimension has a finite extent, $L_5$, and 3-branes (or 4D layers) are placed at $y=0$ and $y=L_5$. Even though the dimensional reduction of all fields in these models is achieved by localization, the mechanism behind localization of gauge fields is still elusive, as charge universality is violated when the usual techniques are employed. In this work, we make a first attempt to investigate the gauge sector of the anisotropic SU(2) lattice gauge theory embedded in a warped metric. The presence of a layered phase might signal a possible non-perturbative way of localizing gauge fields.   
\section{The Mean-Field approach}
The five-dimensional gauge action in the warped background in the continuum is given by
\begin{equation}
S_{AdS_5} = \int d^4x \int dy \Big [ \frac{1}{4g_5^2}F_{\mu\nu}^2 + \frac{1}{2g_5^2}f(y)F_{\mu 5}^2  \Big ] \;; \;\;\;\;\; f(y) = \mbox{e}^{-2ky}
\end{equation}
We call this action $S_{AdS_5}$ as the extra dimension is in a slice of the $AdS_5$ spacetime. Its discretized version imposing an anisotropy is given by
\begin{equation}
S_{AdS_5} = \frac{\beta}{\gamma} \sum_{4D}  \Big (1-\frac{1}{2} \Real \Tr U_{\mu\nu}(n,n_5)\Big) +\beta \gamma \sum_{5D}  \Big (1-\frac{1}{2} \Real \Tr  f(n_5) U_{\mu 5}(n,n_5)\Big)
\end{equation}
where $\gamma$ is the anisotropy parameter and the plaquettes along the usual four dimensions and those extended in the extra dimension are given by Eq.~(\ref{eq.Plaqs})-(\ref{eq.Plaq5}) respectively, where $\mu,\nu=0,1,2,3$
\begin{align}
&U_{\mu\nu}(n,n_5) = U_\mu(n,n_5)U_\nu(n+a_4\hat \mu,n_5) U^\dagger_\mu(n+ a_4\hat \nu,n_5)U^\dagger _\nu(n,n_5) \label{eq.Plaqs} \\ 
&U_{\mu5}(n,n_5) = U_\mu(n,n_5)U_5(n+a_4\hat \mu,n_5) U^\dagger_\mu(n, n_5+ a_5 \hat 5)U^\dagger _5(n,n_5).\label{eq.Plaq5}
\end{align}
The warp factor in our lattice action is anticipated to have an effect on the lattice spacing leading to large finite-size effects. This suggests that the correct investigation of the system using Monte Carlo simulations will be computationally expensive and, as we have no previous studies to guide us to specific regions of parameter space, the first exploration was undertaken using the Mean-Field approximation and specifically employing the saddle-point approach. 
\par 
Following the standard procedure that is described in~\cite{Drouffe:1983fv}, we found the effective action to be
\begin{align}\label{eq:SeffADS}
S_{\rm{eff}} = S_{AdS_5}[V_\mu,V_5] &+ \sum_{n;n_5} \bigg [ \sum_\mu u[H_\mu(n,n_5)] + u_5[H_5(n,n_5)] \nonumber \\
&+ \sum_\alpha h_{\alpha_\mu}(n,n_5) v_{\alpha_\mu}(n,n_5) +  \sum_\alpha h_{\alpha_5}(n,n_5) v_{\alpha_5}(n,n_5) \bigg ]
\end{align} 
where $V$ and $H$ are $2\times 2$ matrices that are used to replace the group-constrained integration measure in the path integral with a flat measure and $v_\alpha$ and $h_\alpha$ are their components after parametrization $(\alpha=0,1,2,3)$. We also define 
\begin{align}
\mathrm{e}^{-u[H_M(n,n_5)]} &= \int_{\SU(2)} {\cal D} U \mbox{e}^{\frac{1}{2} \Real \Tr (UH_M)} 
\end{align} 
which gives 
\be
u\big(H_{M}\big) = -\ln \Big [ \frac{2}{\rho_M}I_1(\rho_M) \Big ] \;\;\; ; \;\;\;
\rho_M = \sqrt{\sum_\alpha (\Real h_{\alpha_M})^2}\;\;\;\; M=\mu, 5.
\ee
where $I_1$ is the modified Bessel function of the first kind of order 1.
\par
Then one usually finds the saddle-point equations and sets the fields to a constant value proportional to the identity matrix. In our case, as the background depends on the extra dimension, there is a mean-field value for each point along the extra dimension. This extra-dimensional dependence of the mean fields was also seen in the construction of the SU(2) theory in an orbifold~\cite{Irges:2012ih} and for our convenience in first-order correction calculations, we made a scale transformation of the fields so that the $AdS_5$ action in Eq.~(\ref{eq:SeffADS}) will look like the flat SU(2) gauge action, i.e. without the factor $f(n_5)$ in front of the extra-dimensional plaquettes. The scaling of the fields is done only on the fields that involve the extra dimension, whereas the fields in the usual four dimensions remain the same
\begin{align} \label{eq:ReDfnv}
&V_\mu(n,n_5) = V'_\mu(n,n_5) \Rightarrow V_{\mu \nu}(n,n_5)= V'_{\mu \nu}(n,n_5) \nonumber \\ 
&V_\mu(n,n_5) = \sqrt{f(n_5)} V'_\mu(n,n_5) \Rightarrow V_{\mu 5}(n,n_5) = f(n_5) V'_{\mu 5}(n,n_5).
\end{align}
Looking at the effective action in Eq.~(\ref{eq:SeffADS}) we see that we also need to rescale $H_5$ as 
\begin{equation}  \label{eq:ReDfnh}
H_5(n,n_5) = \frac{1}{\sqrt{f(n_5)}}H'_5(n,n_5)
\end{equation}
so that we get 
\begin{equation}
h_{\alpha_5}(n,n_5) v_{\alpha 5}(n,n_5) = h'_{\alpha_5}(n,n_5) v'_{\alpha 5}(n,n_5).
\end{equation}
Rescaling the external field in the fifth dimension though, changes the term $u_5[H_5(n,n_5)]$ that becomes
\begin{align}
\mathrm{e}^{-u[H_5'(n,n_5)]}&=\int_{\SU(2)} {\cal D} U \mbox{e}^{\frac{1}{2}\sqrt{f(n_5)} \Real \Tr (U H_5)}.
\end{align} 
The extra factor that involves the warp factor does not affect the nature of the group integral so it can be evaluated as usual using character expansions which results in
\begin{equation}\label{eq:u5}
u_5(H_5') = -\ln \bigg ( \frac{2}{\rho_{5}(n_5) \sqrt{f(n_5)}}I_1\big (\rho_5(n_5)\sqrt{f(n_5)}\big ) \bigg)
\end{equation}
where
 \begin{equation}
 \rho_5(n_5) = \sqrt{\big [ \mbox{Re}(h_{5_0}(n_5)) \big ]^2 + \sum_A\big [ \mbox{Re}(h_{5_A}(n_5))\big ]^2}.
 \end{equation}
Next the saddle-point solutions are determined by setting the fields to a background value which, in contrast to the flat case, has an extra-dimensional dependence, i.e.
\begin{align}
&V_\mu(n,n_5) = \bar v_4(n_5)\mathbb{1}  \;\;\;\;\;\;\;\;\;\; H_\mu(n,n_5) = \bar h_4(n_5) \mathbb{1} \nonumber \\ 
&V_5(n,n_5) = \bar v_5(n_5) \mathbb{1} \;\;\;\;\;\;\;\;\; \;H_5(n,n_5) = \bar h_5(n_5) \mathbb{1}.
\end{align}
This leads to the saddle-point equations given by
\begin{align}\label{eq:SaddlePointEqns}
&\bar v_4(n_5) = \frac{I_2(\bar h_4(n_5))}{I_1(\bar h_4(n_5))} \nonumber \\ 
&\bar v_5 (n_5) = \frac{I_2(\sqrt{f(n_5)}\bar h_5(n_5))}{I_1(\sqrt{f(n_5)}\bar h_5(n_5))} \nonumber \\ 
&\bar h_4(n_5) = 6 \frac{\beta}{\gamma} \bar v_4^3(n_5) + \beta \gamma \bar v_5^2 (n_5) \bar v_4(n_5+a_5) + \beta \gamma \bar v_5^2(n_5-a_5) \bar v_4(n_5-a_5)  \nonumber \\
&\bar h_5(n_5) = 8 \beta \gamma \bar v_5(n_5) \bar v_4(n_5) \bar v_4(n_5+a_5). 
\end{align} 
\section{The phase-diagram}
The first thing we did was to investigate the phase diagram. We made a specific choice of boundary conditions, where we reflected the system in the negative $n_5$ direction and then repeated the system periodically. We call this Periodic Boundary Conditions (PBC) and we have checked with other choices that the system in the middle of the fifth dimension is not affected by the boundary conditions. Then we solved the coupled equations as given in Eq.~(\ref{eq:SaddlePointEqns}) and for each layer (i.e. each $n_5$) we identified three phases according to the following:
\begin{itemize}
\setlength\itemsep{-.5mm}
\item{$v_4(n_5) = 0$, $v_5(n_5) = 0$ Strong-coupling phase~(S) }
\item{$v_4(n_5) \neq 0$, $v_5(n_5) \neq 0$ Deconfining phase~(D)} 
\item{$v_4(n_5) \neq 0$, $v_5(n_5) = 0$ Layered phase~(L)}.
\end{itemize}
We computed the free energy at first order, in an analogous way to~\cite{Irges:2012ih}, to check the stability of the critical points and, as far as we could check, those presented in this phase diagram are stable. 
\par
\begin{figure}[!ht]
\centering
\includegraphics[scale=0.75]{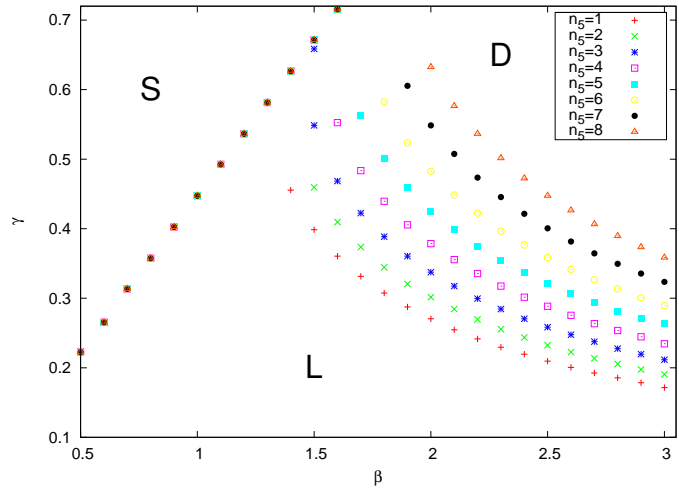}
\caption{The phase diagram obtained for each layer for fixed $k=0.10$. We observe three phases, the confining(S), the deconfining(D) and the layered(L). However, there is a new phase that appears, the mixed phase, in which some of the layers are in the layered phase and some are in the deconfining phase. The width of the mixed phase increases with increasing $k$. }
\label{fig:PhaseDiagram_k010}
\end{figure}

We chose to keep the curvature fixed to the value $k=0.10$ and the lattice size in the positive $n_5$ direction to be $N_5=8$. The layers in the negative $n_5$ direction were matched with layers in the positive $n_5$ direction and thus we consider only the latter in the phase diagram given below. Even though the transition to the confining phase seems to happen at the same point for all layers, we observe that each layer goes from a deconfining phase to the layered phase at different values of $(\beta,\gamma)$. Therefore, we observe an extra phase, \emph{a mixed phase}, where some layers are in the weak-coupling phase and some are in the layered one. This can be seen in Fig.~\ref{fig:PhaseDiagram_k010} as the phase between the orange and the red points. 
\section{The static potential}
\begin{figure}[!ht] 
\centering
\includegraphics[scale=0.32]{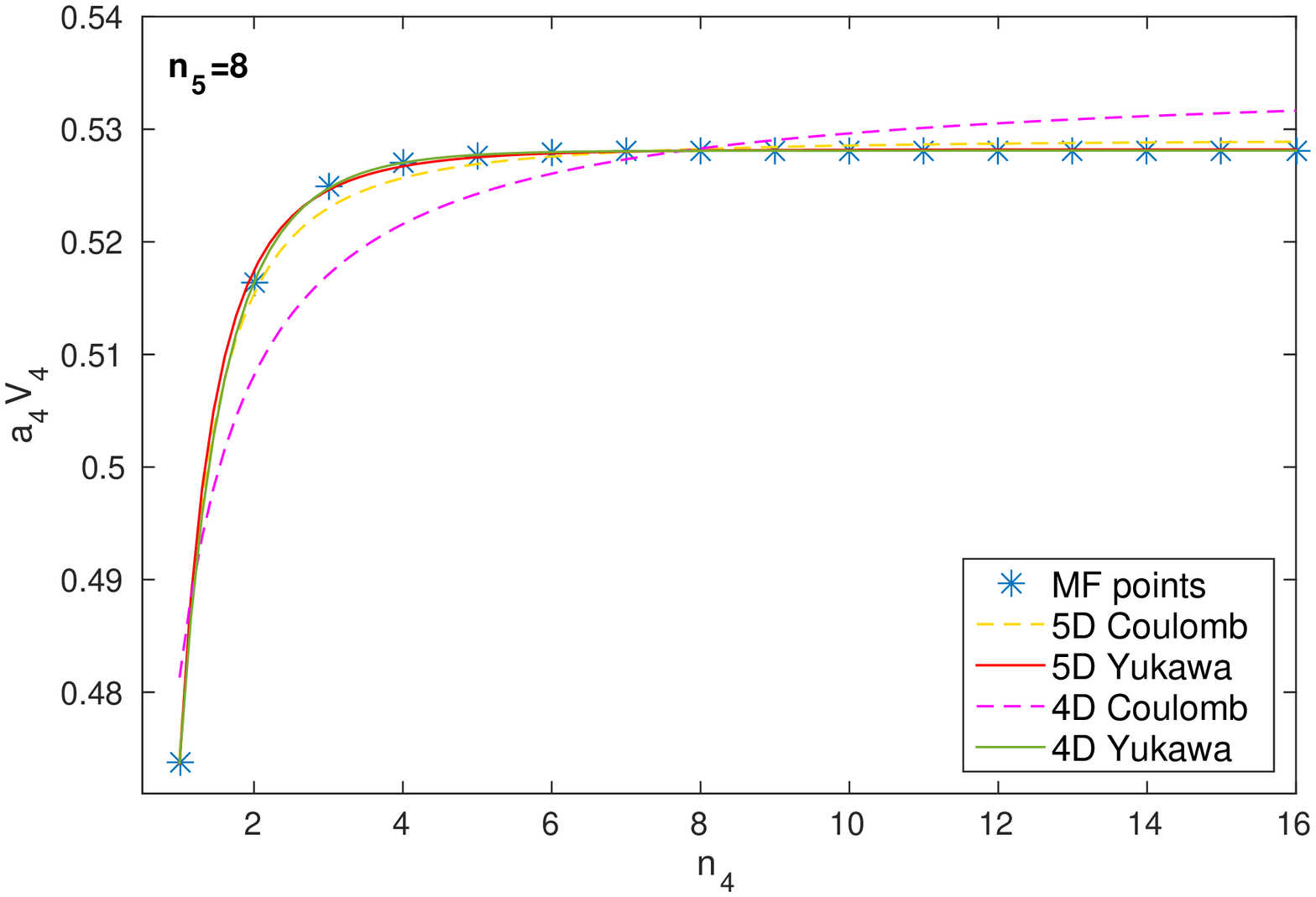}
\includegraphics[scale=0.36]{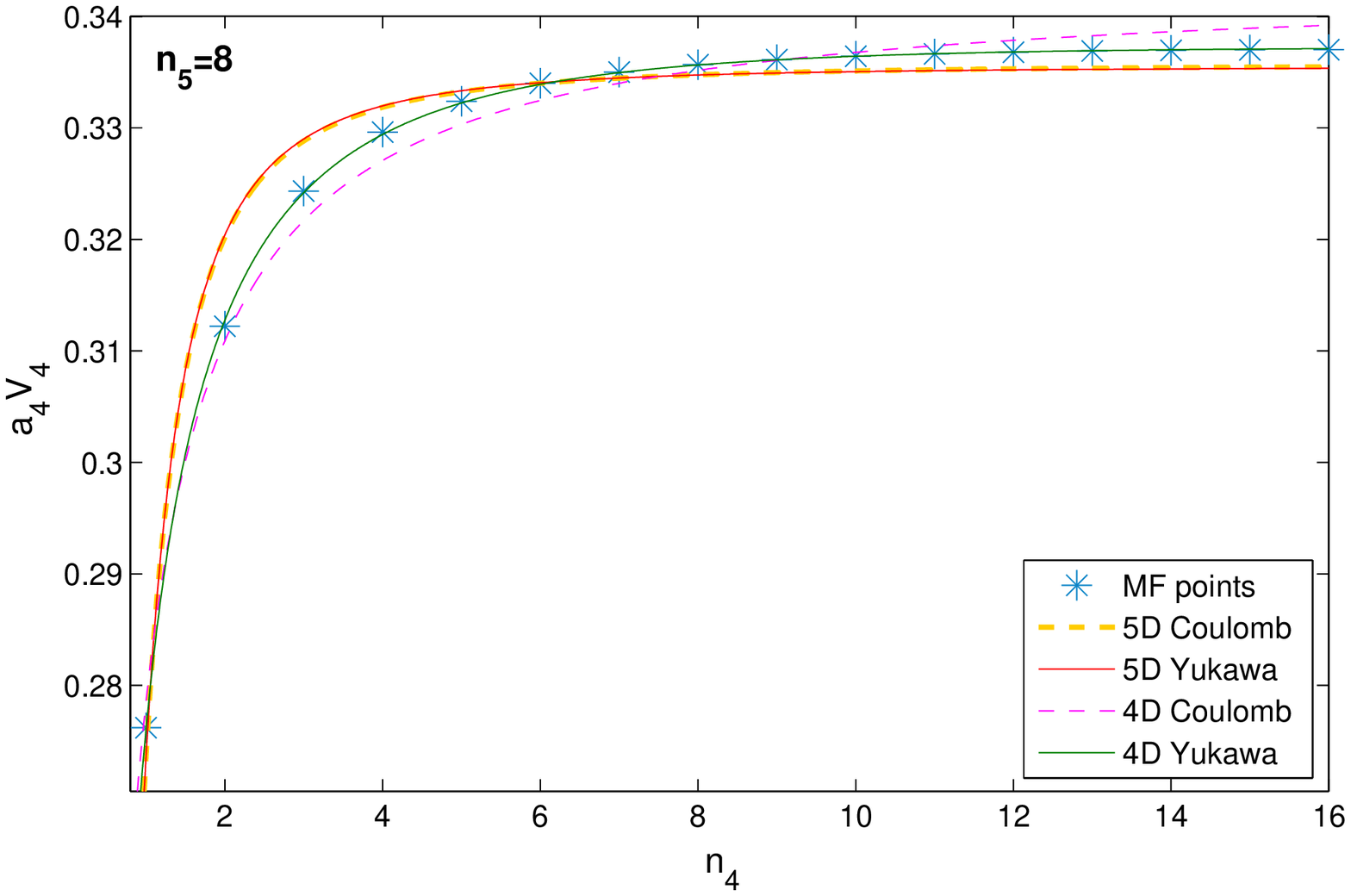}
\caption{Fits to the static potential of the last layer $n_5=8$ using various potential forms for lattice sizes of $T=L=32, N_5=8$ for two different parameter space points: $\beta=2.50$, $\gamma=1.00, k=0.10$~(left) and $\beta=2.30$, $\gamma=0.505, k=0.10$~(right).}
\label{fig:Potential_fits}
\end{figure}

As the main focus of our work is to find evidence of localization of gauge fields, we measured the static potential for each layer at two points in parameter space to investigate its form. Keeping the value $k=0.10$ fixed and a lattice size of $T=L=32,N_5=8$, we chose values of $(\beta,\gamma)$ by inspecting the phase diagram of Fig.~\ref{fig:PhaseDiagram_k010}. The first one was $(2.50,1.00)$ which is away from any phase transition and deep into the deconfining phase. We fitted the mean-field potential points to four different forms: 4D Coulomb, 4D Yukawa, 5D Coulomb and 5D Yukawa. Unfortunately, we could not unambiguously distinguish the form of the potential, as only the 4D Coulomb potential could be excluded, while the rest appeared to be good fits to the potential in all 8 layers. For the last few layers, both 4D and 5D Yukawa forms fitted the MF points well. Fits to all forms of the potential for the last layer $n_5=8$ can be seen in Fig.~\ref{fig:Potential_fits}.
\par
The second point considered was $(2.30,0.505)$, which is close to the transition from the deconfining to the mixed phase. This potential behaves as a 4D Yukawa one for all layers. Also, starting from the first layer, $n_5=1$, the 5D Yukawa and Coulombic potentials also fitted quite well. At larger values of $n_5$, the fits to these forms lose their goodness so, at least for the last layers, we can say with confidence that the potential behaves as a 4D Yukawa one (Fig.~\ref{fig:Potential_fits}). 
\par
\begin{figure}[!ht]
\centering
\includegraphics[scale=0.60]{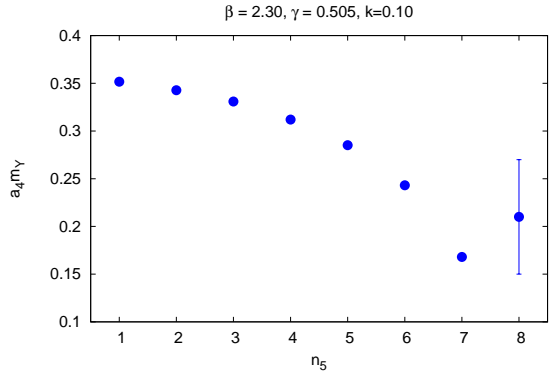}
\caption{The infinite-volume Yukawa mass in lattice spacing units on each 4D layer for $\beta=2.30, \gamma=0.505, k=0.10, N_5=8$ as found by finite-size scaling analysis using lattice sizes of $T=L=24,32,48,100$. All error bars are tiny except for the last layer.}
\label{fig:a4mY_b2300_g0505_k020}
\end{figure}

All the above provide preliminary evidence that, as a Yukawa mass can be obtained, the system close to the transition line is in a 4D Higgs-like phase and not in a Coulombic phase. To check that the Yukawa mass is not the result of the finite extent of our system and remains non-zero in the infinite-volume limit, we performed finite-size scaling on the Yukawa mass and indeed we could get a non-zero value for the infinite-volume Yukawa mass on each layer, as shown in Fig.~\ref{fig:a4mY_b2300_g0505_k020}. This further supports our suspicion that the system is in a Higgs-like phase and not in a Coulombic phase.
\section{Conclusions and Future work }
The mean-field calculations of the static potential show the existence of a Yukawa mass that suggests the presence of a 4D Higgs-like phase close to the line of transition in the phase diagram. This phase suggests that some symmetry breaking may be happening, which is not enforced by imposing certain boundary conditions as done in previous investigations~\cite{Irges:2012ih,Alberti:2015pha}. The only modification in our system from the flat case, where the Higgs-like phase is absent, is the introduction of the curvature along the transverse direction. Thus, we tentatively conclude that the warping breaks the symmetry everywhere in the deconfining phase giving a Higgs-like phase there. This was not clear from the form of the potential away from the transition line, but was not excluded either. So further studies are necessary in order to clarify the nature of the phase in the weak-coupling regime. 
\par 
We return to the question that motivated this project, i.e. whether there is a dimensionally reduced phase close to the layered phase. 
If there is a 5D Higgs-like phase away from the transition line, then we might have dimensional reduction via localization, analogous to the one found in~\cite{Alberti:2015pha}, where they explicitly broke the symmetry using the orbifold. If not, then the system, due to the warping, behaves as a four-dimensional one everywhere outside the strong-coupling phase. It is noteworthy that we have used a small extent of lattice points along the extra dimension, which restricts the region of the mixed phase to a small width. It appears likely that the pure deconfining phase is a finite-size effect of the fifth direction, and the infinite system is actually in a 4D Higgs-like phase everywhere in the weak-coupling regime.
 \par
There are a number of open questions that still need to be resolved by further work. One is whether this Higgs-like phase is physical, a lattice artefact, or a fake result of the Mean-Field approximation. Also, nothing can be said about the layered phase at the moment. Studies using Monte Carlo simulations are expected to show the true behaviour in this phase, which might be 4D Higgs-like, or Coulombic. All in all, our tentative conclusion that there is a 4D Higgs-like phase motivates a range of further tests and explorations, especially with numerical simulations, to clarify the effect of warping.
\section*{Acknowledgments}
E.L. is supported by an STFC studentship. We thank F. Knechtli for the fruitful discussions during the conference.

\end{document}